\documentclass[]{elsart}
\usepackage{graphics}
\usepackage{graphicx}
\usepackage{epsfig}
\usepackage{amssymb}
\begin{document}
\begin{frontmatter}
\title{Correlation Networks Among Currencies}


\author{\small{Takayuki Mizuno$^1$$^{a}$}},
\author{\small{Hideki Takayasu$^{b}$}},
\author{\small{Misako Takayasu$^{a}$}}

\address{$^{a}$Department of Computational Intelligence and Systems Science, Interdisciplinary Graduate School of Science and Engineering, Tokyo Institute of Technology, G3-52 4259 Nagatuta-cho, Yokohama 226-8502, Japan}
\address{$^{b}$Sony Computer Science Laboratories Inc., 3-14-13 Higashigotanda, Shinagawa-ku, Tokyo 141-0022, Japan}

\thanks{Corresponding author.\\
{\it E-mail address:}\/ mizuno@smp.dis.titech.ac.jp (T.Mizuno)}

\begin{abstract}
By analyzing the foreign exchange market data of various currencies, we 
derive a hierarchical taxonomy of currencies constructing minimal-spanning 
trees. Clustered structure of the currencies and the key currency in each 
cluster are found. The clusters match nicely with the geographical regions 
of corresponding countries in the world such as Asia or East Europe, the key 
currencies are generally given by major economic countries as expected.
\end{abstract}


\end{frontmatter}

\textbf{Introduction} \\
A value of currency is expected to reflect the whole economic status of the 
country, and a foreign exchange rate is considered to be a measure of 
economic balance of the two countries. In the real world there are several 
economic blocks such as Asia, but it is not clarified whether such economic 
blocks affect the foreign exchange rate fluctuations or not. From the 
viewpoint of physics, the foreign exchange market is a typical open system 
having interactions with all information around the world including price 
changes of other markets. Also, the mean transaction intervals of foreign 
exchange markets are typically about 10 seconds, and it is not clear how the 
market correlates with the huge scale information of a whole country or the 
economic blocks. In order to empirically establish the relations between 
microscopic market fluctuations and macroscopic economic states, it is 
important to investigate the interaction of currency rates in the high 
precision data of foreign exchange markets. 

The correlations among market prices have been analyzed intensively for 
stock prices by using minimal-spanning trees or self-organizing maps [1] [2] 
[3] [4] [5]. The interaction among stocks is expected to be caused by 
information flow, and direction of the information flow has been 
investigated from a cross-correlation function with a time shift [6][7][8]. 
L. Kullmann, et al. and J. Kertesz, et al. introduced a directed network 
among companies for the stocks [7][8]. We observe the interaction among 
foreign exchange markets using minimal-spanning tree.

We construct a currency minimal-spanning tree by defining correlation among 
foreign exchange rates as the distance. The minimal-spanning tree is a kind 
of currency map and is helpful for constructing a stable portfolio of the 
foreign exchange rates. We use correlation coefficient of daily difference 
of the logarithm rate in order to detect the topological arrangement of the 
currencies. The correlation coefficient is computed between all the possible 
pairs of rates in a given time period. We classify the currencies on the 
minimal-spanning tree according to the correlation coefficients, and find 
key currencies in each cluster. We analyze 26 currencies and 3 metals from 
January `99 up to December `03 provided by Exchange Rate Service [9].

\textbf{Method of hierarchical taxonomy of currencies} \\
We introduce a method of hierarchical taxonomy of currencies. We first 
define correlation function between a pair of foreign exchange rates in 
order to quantify synchronization between the currencies. We focus on a 
daily rate change $dP_i (t)$ defined as

\begin{equation}
\label{eq1}
dP_i (t) \equiv \log P_i (t + 1\mbox{day}) - \log P_i (t),
\end{equation}

where $P_i (t)$ is the rate $i$ at the time $t$. Using the rate change, 
correlation coefficient between a pair of the rates can be calculated by 
cross-correlation function as

\begin{equation}
\label{eq2}
C_{ij} = \frac{\left\langle {dP_i \cdot dP_j } \right\rangle - \left\langle 
{dP_i } \right\rangle \left\langle {dP_j } \right\rangle }{\sqrt 
{(\left\langle {dP_i^2 } \right\rangle - \left\langle {dP_i } \right\rangle 
^2)(\left\langle {dP_j^2 } \right\rangle - \left\langle {dP_j } 
\right\rangle ^2)} },
\end{equation}

where $\left\langle {dP_i } \right\rangle $ represents the statistical 
average of $dP_i (t)$ for a given time. The correlation coefficient $C_{ij} 
$ has values ranging from --1 to 1. 

We get $n\times n$ matrix of $C_{ij} $ by calculating the cross-correlation 
function for all combinations among the given rates when $n$ kind of foreign 
exchange rates are given. It is clear that the matrix has symmetry $C_{ij} = 
C_{ji} $ with $C_{ii} = 1$ from the definition of Eq.(\ref{eq2}). We apply the 
correlation matrix to construct a currency minimal-spanning tree (MST), and 
can intuitively understand network among the foreign exchange rates using 
the MST. The MST forms taxonomy for a topological space of the $n$ rates. The 
MST is a tree having $n - 1$ edges that minimize the sum of the edge 
distances in a connected weighted graph of the $n$ rates. The edge distances 
satisfy the following three axioms of a Euclidean distance: (i) $d_{ij} = 0$ 
if and only if $i = j$, (ii) $d_{ij} = d_{ji} $, (iii) $d_{ij} \le d_{ik} + 
d_{kj} $. Here, $d_{ij} $ expresses a distance for a pair of the rate $i$ 
and the rate $j$. We need Euclidean distances between the rates in order to 
construct the MST. However, the correlation coefficient $C_{ij} $ does not 
satisfy the axioms. We can convert the correlation coefficient by 
appropriate functions so that the axioms can be applied [1]. One of the 
appropriate functions is

\begin{equation}
\label{eq3}
d_{ij} = \sqrt {2(1 - C_{ij} )} ,
\end{equation}

where $d_{ij} $ is a distance for a pair of the rate $i$ and the rate $j$.

We construct a MST for the $n$ rates using $n\times n$ matrix of $d_{ij} $. One 
of methods which construct MST is called Kruskal's algorithm [10][11]. The 
Kruskal's algorithm is a simple method consisting of the following steps: In 
the first step we choose a pair of rates with nearest distance and connect 
with a line proportional to the distance. In the second step we also connect 
a pair with the 2$^{nd}$ nearest distance. In the third step we also connect 
the nearest pair that is not connected by the same tree. We repeat the third 
step until all the given rates are connected in one tree. Finally, we 
achieve a connected graph without cycles. The graph is a MST linking the $n$ 
rates.

We introduce maximal distance $\widehat{d}_{ij} $ between two successive 
rates encountered when moving form the starting rate $i$ to the ending rate 
$j$ over the shortest part of the MST connecting the two rates. For example, 
the distance $\widehat{d}_{ad} $ is $d_{bc} $ when the MST is given as

\begin{center}
$a$ -- $b$ --- $c$ -- $d,$
\end{center}

where $d_{bc} \ge \max \left\{ {d_{ab} ,d_{cd} } \right\}$. The distance 
$\widehat{d}_{ij} $ satisfies axioms of Euclidean distance and a following 
ultrametric inequality with a condition stronger than the axiom (iii) $\hat 
{d}_{ij} \le \hat {d}_{ik} + \hat {d}_{kj} $ [12],

\begin{equation}
\label{eq4}
\widehat{d}_{ij} \le \max \left\{ {\widehat{d}_{ik} ,\widehat{d}_{kj} } 
\right\}.
\end{equation}

The distance $\widehat{d}_{ij} $ is called Subdominant ultrametric distances 
[10][13]. A space connected by the distances provides a well defined 
topological arrangement that has associated with a unique indexed hierarchy. 
For a set of foreign exchange rates, we describe the hierarchy by 
constructing MST. The result will be elaborated in the next section.

\newpage
\textbf{Correlation networks among currencies} \\
The traders in a foreign exchange market are always observing many other 
markets. Among them, they pay a special attention to the currencies of the 
countries which economically influences the country using the currency they 
are trading. For example, traders of Swiss Franc pay attention to Euro. 
Therefore, correlation between CHF(Swiss Franc)/USD and EUR(Euro)/USD is 
strong and there is a time delay of order less than a minute between the two 
rates because changes of EUR/USD feed back to CHF/USD of the future [6]. We 
investigated 26 currencies and 3 metals in New York market from January `99 
to December `03 as listed in Table.1. Probability density distributions of 
correlation coefficients among the currencies and the metals measured by 
USD(United States Dollar) for each year are shown in Fig.1. Here, the 
correlation coefficients are calculated from logarithm rate changes. The 
nontrivial various correlations are found for each year. 

We clarify the market networks using MST. We first analyze the currencies 
and the metals measured by USD. The MST is constructed by the Kruskal's 
algorithm using database of changes of the foreign exchange rates and the 
metal prices. We show the MST and an indexed hierarchical tree associated 
with the MST in Fig.2(a) and (b). We focus on EUR in Fig.2(a). Neighbors of 
the EUR are European currencies, such as Swiss Franc, Hungarian Forint, and 
Norwegian Krone. Other currencies also connect with the currencies of 
geographically close countries in the MST. From these results we notice that 
the currencies and the metals cluster with East Europe, West Europe, 
Oceania, South Amelica, Asia, and metal. 

We can find more clearly these clustered structures by observing the indexed 
hierarchical tree in conjunction with the MST in Fig.2(b). In the right side 
of the indexed hierarchical tree, the West Europe cluster (SEK-Swedish 
Krona, NOK-Norwegian Krone, CHF, EUR) and the East Europe cluster 
(SKK-Slovakian Koruna, CZK-Czech Koruna, HUF-Hungarian Forint) connect 
between EUR and HUF form Europe cluster. The Europe cluster also connects 
GBP-British Pound and PLZ-Polish Zloty. We can also clearly find the Oceania 
cluster (AUD-Australian Dollar, NZD-New Zealand Dollar), the Asian cluster 
(IDR-Indonesian Rupee, JPY-Japanese Yen, SGD-Singapore Dollar, THB-Thai 
Baht), the South American cluster (BRR-Brazilian Real, CLP-Chilean Peso, 
MXP-Mexican Peso), and the metal cluster (Au-Gold, Ag-Silver, Pt-Platinum). 
The key currencies which connect with some currencies are EUR, HUF, AUD, JPY 
and MXP in the clusters. Therefore, the clusters match nicely with the 
geographical regions of corresponding countries in the world, and the key 
currencies are generally given by major economic countries.

We next investigate relations between USD and other currencies. The metal 
cluster is almost independent of the currency clusters in Fig.2(a) and (b). 
Especially, the platinum most loosely connects with the currencies in the 
metal cluster. We focus on the platinum with few influences to the 
currencies, and construct a MST using the currencies and the metals measured 
by the platinum. Fig.3(a) and (b) show the MST and an indexed hierarchical 
tree associated with the MST. Unlike Fig.2(a), one currency has big 
influence to other many currencies in Fig.3(a). The currency is USD as 
naturally expected; namely, USD has substantial weight in the global world 
economy. There are some European currencies centered around the EUR in the 
left side of Fig.3(a). The currencies (EUR, CHF, NOK, SEK, GBP) form the 
hierarchical tree which does not contain USD in Fig.3(b). Only European 
currencies are influenced by EUR rather than USD.

\textbf{Discussion} \\
Each country's currency influences currencies of neighboring countries. We 
showed correlation networks among currencies by using MST. We found some 
clusters in the correlation networks. The clusters match nicely with the 
geographical regions of corresponding countries in the world. The key 
currencies are generally given by major economic countries in the clusters. 
It was confirmed that USD is virtually the standard global currency because 
especially USD has big influence to other currencies. Therefore, minor 
currency depends on USD and the key currency of the region where it belongs. 
Traders generally handle the job of only one foreign exchange rate in 
exchange markets. Therefore, the dependence among currencies means that the 
traders are always observing not only the market they trade in, but also 
markets of USD and the key currencies. They feed back the changes of the 
currencies in their own trading.

In financial engineering, many models independently describe a foreign 
exchange rate, assuming that interactions among different currencies can be 
processed by random noises of the exchange rate based on the concepts of a 
mean field approximation. The theory should be improved for minor 
currencies. 

We expect that the hierarchical taxonomy of currencies is helpful for the 
improvement. We finally discuss the correlation networks from a standpoint 
of monetary systems. In international trades between two traders, one of the 
both traders has to exchange currency in an exchange market except when the 
both traders are using the same currency. Therefore, there are exchange 
risks, such as exchange fee, in the international trades. Governments are 
interested in regional currency without the exchange risks such as the Euro 
because they want to invigorate international trades in a region. When 
introducing the regional currency, we have to determine basket ratio of the 
regional currency in consideration of economical dependency among the 
countries. The economical dependency can be clarified using our theory of 
the correlation networks. Therefore, we expect that the correlation networks 
are helpful in future monetary system.

\textbf{Acknowledgement} \\
T. Mizuno is supported by Research Fellowships of the Japan Society for the 
Promotion of Science for Young Scientists.




\newpage
Table.1 The set of Daily data for 26 currencies and 3 metal

\begin{tabular}{ll}
\textbf{Code} & \textbf{Currency} \\
AUD  & Australian Dollar \\
BRR  & Brazilian Real \\
GBP & British Pound \\
CAD & Canadian Dollar \\
CLP & Chilean Peso \\
COP & Colombian Peso \\
CZK & Czech Koruna \\
EUR & Euro \\
HUF & Hungarian Forint \\
IDR & Indonesian Rupiah \\
JPY & Japanese Yen \\
MXP & Mexican Peso \\
NZD & New Zesland Dollar \\
NOK & Norwegian Kroner \\
PEN & Peruvian New Sole \\
PHP & Philippines Peso \\
PLZ & Polish Zloty \\
RUR & Russian Ruble \\
SGD & Singapre Dollar \\
SKK & Slovakian Koruna \\
ZAR & South African Rand \\
KRW & South Korean Won \\
SEK & Swedish Krona \\
CHF & Swiss Franc \\
THB & Thai Baht \\
USD & U.S. Dollar \\
Au & Ounce of Gold in New York market \\
Ag & Ounce of Silver in New York market \\
Pt & Ounce of Platinum in New York market \\

\end{tabular}

\newpage

Fig.1 PDF of correlation coefficient for currencies measured by US dollar 
for every year.

Fig.2 (a): MST for the currencies and the metals measured by USD. (b): 
Indexed hierarchical tree obtained for the currencies and the metals.

Fig.3 (a): MST for the currencies and the metals measured by ounce of 
platinum. (b): Indexed hierarchical tree obtained for the currencies and the 
metals.

\end{document}